\documentclass[osajnl2,josaa,twocolumn,amsmath]{revtex4}  
\usepackage{graphicx}
\usepackage{epstopdf}
\usepackage{psfrag}
\newcommand{\boldmatrix}[1]{\boldsymbol{\mathcal{#1}}}
\DeclareMathOperator{\EF}{\mathit{F}}
\DeclareMathOperator{\ER}{\widetilde{\mathit{F}}}
\begin{document}
\title{Shifted-elementary-mode representation for partially coherent
vectorial fields}

\author{Jani Tervo}
\author{Jari Turunen}
\author{Pasi Vahimaa}
\affiliation{University of Eastern Finland, Department of Physics and Mathematics, P.O. Box 111, FI-80101 Joensuu, Finland}

\author{Frank Wyrowski}
\affiliation{Friedrich Schiller University of Jena, Department of
Applied Physics, D-07745 Jena, Germany}

\begin{abstract}
A representation of partially spatially coherent and partially
polarized stationary electromagnetic fields is given in terms of
mutually uncorrelated, transversely shifted, fully coherent and
polarized elementary electric-field modes. This representation
allows one to propagate non-paraxial partially coherent vector
fields using techniques for spatially fully coherent fields, which
are numerically far more efficient than methods for propagating correlation
functions. A procedure is given to determine the elementary modes from
the radiant intensity and the far-zone polarization properties of
the entire field. The method is applied to quasi\-homogeneous fields
with rotationally symmetric $\cos^n\theta$ radiant intensity distributions
($\theta$ being the diffraction angle with respect to the optical axis and
$n$ an integer). This is an adequate model for fields emitted by, e.g.,
many light-emitting diodes.
\end{abstract}


\maketitle

\section{Introduction}

There is a growing demand to include partial spatial coherence in optical design,
not only because of classical areas of application such as microscopy and projection
lithography, but also because of the increasing importance of partially coherent solid-state
sources such as multimode lasers (including excimers) and LEDs.
Methods based on ray optics can not adequately describe all of the relevant issues related to
coherence and polarization, which are intimately connected in electromagnetic coherence theory.
Thus wave-optical methods are required, which must be capable of dealing with non-paraxial fields
and systems that may contain also micro- and nano\-structures in
objects and interfaces. To this end, one needs computationally efficient physical-optics-based
representations of spatially partially coherent electromagnetic fields.

Apart from some specific models that allow analytic solutions, propagating spatially partially coherent
light even in free space is a formidable numerical problem involving four-dimensional integrals~\cite{Mandel}.
The dimensionality of the propagation integrals can be decreased to two (for planar sources) if the partially coherent field is represented as an incoherent superposition of fully spatially coherent fields. The classical way to do this is the coherent-mode expansion of the cross-spectral density (CSD) function by means of Mercer's expansion~\cite{Wolf82}, which has recently been extended to electromagnetic fields~\cite{Gori03,Tervo04}. In this representation the coherent modes are uniquely defined by the CSD through a Fredholm integral equation; they form a complete and orthonormal set, in which the effective number of modes $N$ increases as the degree of coherence of the field is reduced~\cite{Gori80a,Gori80b,Starikov1,Starikov2}. Thus, in free-space propagation, the original four-dimensional integral for CSD is replaced by $N$ two-dimensional integrals for the coherent modes. In light-matter interaction analysis one solves the diffraction or scattering problem for $N$ coherent fields of different functional forms~\cite{Huttunen,Vahimaa97}.

There is also an alternative representation of a partially coherent field in terms of uncorrelated, fully coherent fields~\cite{Gori78,Gori80,Vahimaa,Gori07,Turunen08}. Here all coherent fields or `elementary modes' are of identical functional form but spatially (or angularly) shifted with respect to each other and weighted by a function determined by the CSD. Unlike the Mercer expansion, the shifted-elementary-mode representation is applicable only to a specific class of genuine CSDs~\cite{Gori07}. However, this class contains many of the fields of practical significance in optical design, including all quasihomogeneous fields (LEDs, excimers, illumination in microscopy and projection lithography, etc.). Since the elementary modes are identical, only a single 2D integral needs to be evaluated in free-space propagation problems. In interaction problems one needs to scan the elementary mode across the object and perform a set of independent diffraction calculations for coherent light. Typically the elementary mode has a smooth functional form, at least compared to the higher-order modes in the Mercer expansion, and is therefore easy to propagate numerically. Another advantage of the shifted-mode model is that there is no need for numerical solution of the Fredholm integral equation:  the elementary mode and the associated weight function can be determined, e.g., from far-zone properties of the field~\cite{Vahimaa}.

In this paper we generalize the scalar shifted-elementary-mode representation to the vectorial case, which is necessary to adequately model partially spatially coherent, partially polarized sources. There are two major reasons why such a generalization is necessary: first, non-paraxial fields can not be adequately described by a scalar model and, second, optical components in the system can modify the polarization state of the field. It turns out that the vectorial nature of the field does not fundamentally complicate the numerical procedure. Two elementary modes are needed to specify the state of polarization, but also in the electromagnetic case the modes and their weight functions can be determined from far-zone properties of the field.

We begin the discussion by briefly reviewing the scalar model in Sect.~\ref{s:scalar} to establish the notation and to simplify the interpretation of the main results. The extension to the electromagnetic case is outlined in Sect.~\ref{s:emext} and the rigorous mathematical formulation is presented Sections \ref{s:far} and \ref{s:elementary}. The important special case of rotationally symmetric and quasihomogeneous fields is discussed in Sections \ref{s:rotsym} and ~\ref{s:quasihom}, respectively. Some numerical results are provided in Sect.~\ref{s:cosine}. In Sect.~\ref{s:LED} we apply the model to a simple LED geometry. Finally, issues such as the measurements required to determine the elementary modes and their weight functions are discussed in Sect.~\ref{s:discussion}.

\section{The scalar model}
\label{s:scalar}

\begin{figure}
\psfrag{a}{(a)}\psfrag{b}{(b)}
\psfrag{x}{$x$}\psfrag{y}{$y$}\psfrag{z}{$z$}
\psfrag{4}{$k_x$}\psfrag{q}{$k_y$}\psfrag{2}{$k_z$}
\psfrag{3}{$\psi$}\psfrag{r}{$r$}\psfrag{2}{$k_z$}
\psfrag{w}{$\kappa$}\psfrag{h}{$\mathbf{\hat{x}}$}\psfrag{o}{$\mathbf{\hat{z}}$}
\psfrag{e}{$\rho$}\psfrag{k}{$k$}\psfrag{t}{$\theta$}
\psfrag{g}{$\phi$}\psfrag{n}{$\mathbf{\hat{s}}$}
\psfrag{l}{$\mathbf{\hat{r}}$} \psfrag{i}{$\mathbf{\hat{\phi}}$}
\psfrag{j}{$\boldsymbol{\hat{\rho}}$}
\psfrag{6}{$\boldsymbol{\hat{\theta}}$}
\psfrag{8}{$\boldsymbol{\hat{\kappa}}$}
\psfrag{5}{$\boldsymbol{\hat{\psi}}$} \psfrag{u}{$\mathbf{\hat{k}}$}
\psfrag{v}{$\boldsymbol{\kappa}$}\psfrag{c}{$\mathbf{\hat{y}}$}\psfrag{7}{$\mathbf{k}$}
\psfrag{f}{$\boldsymbol{\rho}$}\psfrag{m}{$\mathbf{r}$}
\psfrag{d}{$\boldsymbol{\sigma}$} \psfrag{0}{$z=0$}
\begin{center}
\includegraphics[width=\columnwidth]{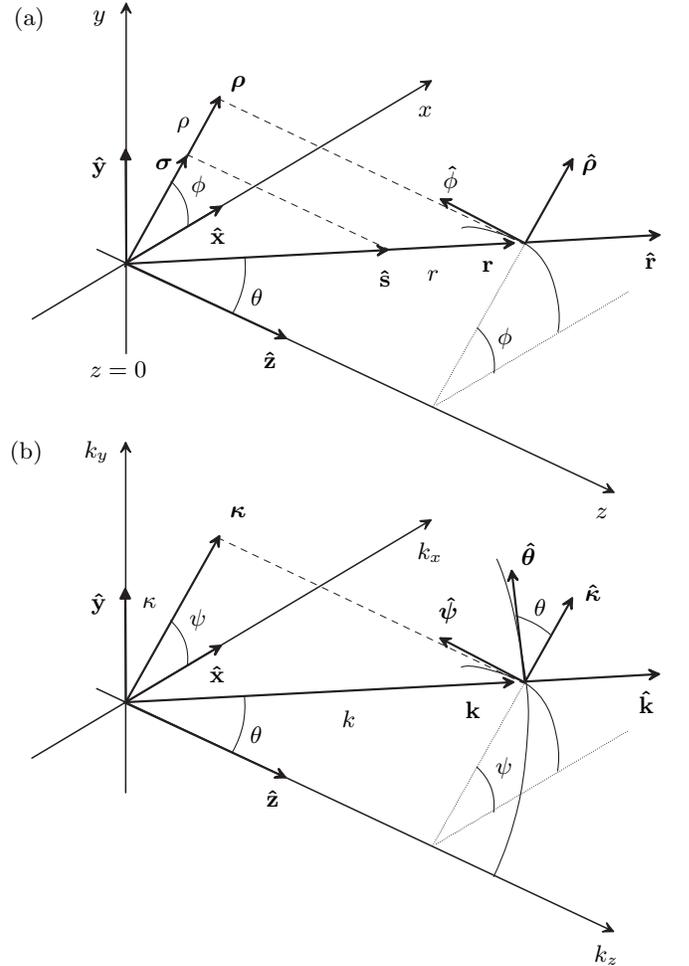}
\end{center}
\caption{Notations used for (a) position and (b) wave-vector
coordinates.} \label{f:notations}
\end{figure}

Using the notations of Fig.~\ref{f:notations}, we may write the well-known relationship~\cite{Mandel} between the cross-spectral density function $W(\mathbf{r}_1,\mathbf{r}_2)$ and the angular correlation function $A(\boldsymbol{\kappa}_1,\boldsymbol{\kappa}_2)$ as (we omit the dependence on the angular frequency $\omega$ for brevity though\-out the paper)
\begin{align}
W(\mathbf{r}_1,\mathbf{r}_2) &=\frac{1}{(2\pi)^4}
\iiiint_{-\infty}^\infty
A(\boldsymbol{\kappa}_1,\boldsymbol{\kappa}_2)\nonumber\\
&\quad\times \exp\left[{\rm
i}(\mathbf{k}_2\cdot\mathbf{r}_2-\mathbf{k}_1^*\cdot\mathbf{r}_1)\right]
\mathrm{d}^2\kappa_1\,\mathrm{d}^2\kappa_2. \label{angspecscalar}
\end{align}
Here the asterisk indicates complex conjugation, $\mathbf{r}_j = \left(x_j,y_j,z_j\right)$ with $j = 1,2$ are position vectors, $\mathbf{k}_j = \left(k_{jx},k_{jy},k_{jz}\right) = \left(\boldsymbol{\kappa}_{j},k_{jz}\right)$ represent wave vectors, and $\boldsymbol{\kappa}_j=\left(k_{jx},k_{j,y}\right)$ are their transverse projections.

Restricting now to the specific class of fields mentioned in the introduction, we assume that the angular correlation
function is of the Schell-model form~\cite{Vahimaa}
\begin{equation}
A(\boldsymbol{\kappa}_1,\boldsymbol{\kappa}_2) = g(\Delta\boldsymbol{\kappa})f^\ast(\boldsymbol{\kappa}_1)f(\boldsymbol{\kappa}_2).
\label{Aschell}
\end{equation}
The radiant intensity of a scalar field is defined as~\cite{Mandel}
\begin{equation}
J(r\hat{\mathbf{s}}) = 2 n \pi^2k^2\cos^2\theta A(k\boldsymbol{\sigma},k\boldsymbol{\sigma}),
\label{Jdefscalar}
\end{equation}
where $n$ is the refractive index of the medium, $\mathbf{\hat{s}}=\mathbf{r}/r$ with $r=\|\mathbf{r}\|$
is a unit direction vector, $\boldsymbol{\sigma}$ is its transverse projection, $\theta$ is the angle between
$\mathbf{\hat{s}}$ and the $z$ axis, and $k=\|\mathbf{k}\|$ is the wave number. Using Eq.~\eqref{Aschell}, we have
\begin{equation}
J(r\hat{\mathbf{s}}) = 2 n \pi^2k^2\cos^2\theta \left|f(k\boldsymbol{\sigma})\right|^2
\label{Jexprscalar}
\end{equation}
Thus the radiant intensity of a Schell-model partially coherent field defined by Eq.~\eqref{Aschell} is the same as that produced by a coherent field with angular spectrum $f(\boldsymbol{\kappa})$.

Let us introduce two-dimensional Fourier-transform relations
\begin{equation}
e(\mathbf{r}) = \frac{1}{(2\pi)^2}\iint_{-\infty}^\infty f(\boldsymbol{\kappa})\exp\left({\rm i}\mathbf{k}\cdot\mathbf{r}\right){\rm d}^2\kappa
\label{IFT-e}
\end{equation}
and
\begin{equation}
p(\boldsymbol{\rho}) = \frac{1}{(2\pi)^2}\iint_{-\infty}^\infty g(\Delta\boldsymbol{\kappa})\exp\left({\rm i}\Delta \boldsymbol{\kappa}\cdot\boldsymbol{\rho}\right){\rm d}^2\Delta\kappa,
\label{IFT-p}
\end{equation}
where $\Delta\boldsymbol{\kappa} = \boldsymbol{\kappa}_2-\boldsymbol{\kappa}_1$ and $\boldsymbol{\rho}=x\mathbf{\hat{x}}+y\mathbf{\hat{y}}$ is the transverse projection of the position vector.
Inserting Eq.~\eqref{Aschell} into Eq.~\eqref{angspecscalar} and using the Fourier representation of $g(\Delta\boldsymbol{\kappa})$ obtained by inverting Eq.~\eqref{IFT-p}, we get the expression
\begin{equation}
W(\mathbf{r}_1,\mathbf{r}_2) = \iint_{-\infty}^\infty p(\boldsymbol{\rho}^\prime)e^\ast(\mathbf{r}_1-\boldsymbol{\rho}^\prime)e(\mathbf{r}_2-\boldsymbol{\rho}^\prime)
{\rm d}^2\rho^\prime
\label{Scalarrep}
\end{equation}
for the CSD~\cite{Vahimaa}. This result applies, in particular, at $z = 0$. Thus $e(\boldsymbol{\rho},0)$ is the coherent source-plane field with angular spectrum $f(\boldsymbol{\kappa})$. The representation in Eq.~\eqref{Scalarrep} expresses the partially coherent field as a weighted linear superposition of spatially shifted but identical fully coherent elementary (scalar) fields $e(\boldsymbol{\rho},0)$.

\section{The electromagnetic extension}
\label{s:emext}

As is evident from Eq.~\eqref{Jdefscalar}, the mathematical form of the elementary field mode $e(\boldsymbol{\rho},0)$ can be determined from the knowledge of the radiant intensity (at least apart from a phase factor, which can in fact be employed to model volume sources~\cite{Turunen08}). In general, the weight function $p(\boldsymbol{\rho}^\prime)$ can be determined from far-field coherence measurements using Eqs.~\eqref{Aschell} and \eqref{IFT-p}, but often there are simpler ways to at least approximate it~\cite{Vahimaa}. This is the case, in particular, if the field is quasi\-homogeneous, i.e., if the coherence area in the source plane is much smaller than source area; in this case the weight function comes out of the integral in Eq.~\eqref{Scalarrep}. When $e(\boldsymbol{\rho},0)$ and $p(\boldsymbol{\rho}^\prime)$ are known, coherent propagation techniques for $e(\boldsymbol{\rho},0)$ and a linear superposition according to Eq.~\eqref{Scalarrep} suffice to propagate the entire spatially partially coherent field.

The extension of the scalar shifted-mode model to the electromagnetic case is not trivial. One might be tempted to use some superpositions of, e.g., locally linearly polarized modes of the scalar functional form to model partially polarized or unpolarized sources. However, such constructions seem hard to justify mathematically. The approach taken here is based on far-field information: in the far zone the field behaves as an outgoing spherical wave, and therefore it has a well-defined local polarization state. We employ this fact to separate the angular correlation tensor, which is the electromagnetic extension of the function $A(\boldsymbol{\kappa}_1,\boldsymbol{\kappa}_2)$, into two orthogonal parts [see Eq.~\eqref{A:StartingPoint} in sect.~\ref{s:far}]. These represent the electromagnetic elementary modes (or polarization modes) in the far field. Then the source-plane modes can be determined by Fourier-transform techniques in analogy with the scalar case.

As a result of the construction process (to be described mathematically in the following sections), we obtain a vectorial shifted-mode expansion for the spatially partially coherent field everywhere in space [see Eqs.~\eqref{FieldSource}, \eqref{Elementary}, and \eqref{ElementaryB} in Sect.~\ref{s:elementary}]. Instead of propagating one fully coherent mode as in the scalar case, we now need to propagate two well-defined fully coherent vectorial field modes, and to form the generalized shifted-mode superposition, to govern the propagation of the electromagnetic spatially partially coherent field. Thus the increase in computational complexity, compared to the scalar case, is essentially fourfold.

\section{Field representation in the far-zone}
\label{s:far}

A statistically stationary random electromagnetic field in
the space--frequency domain is described by a CSD matrix
$\boldmatrix{W}(\mathbf{r}_1,\mathbf{r}_2)$, which may be
expressed as~\cite{Tervo04,Alonso08}
\begin{align}
\boldmatrix{W}(\mathbf{r}_1,\mathbf{r}_2) = \langle
\mathbf{E}^\ast(\mathbf{r}_1)\mathbf{E}^\mathrm{T}(\mathbf{r}_2)\rangle.
\label{Wpqdef}
\end{align}
Here $\mathrm{T}$ indicates the transpose, the brackets denote ensemble
averaging, and the electric-field realizations
$\mathbf{E}(\mathbf{r})$ are understood as appropriate random linear
superpositions of the eigenfunctions of the Fred\-holm integral
equation satisfied by
$\boldmatrix{W}(\mathbf{r}_1,\mathbf{r}_2)$~\cite{Tervo04,Gori03}.
The position-dependent spectral density of the field can be written,
in analogy with scalar theory of partial coherence~\cite{Wolf82}, as
$S(\mathbf{r}) = {\rm tr}\,\boldmatrix{W}(\mathbf{r},\mathbf{r})$,
where tr stands for trace.

The relation between the field at the (secondary) source plane $z=0$
and the far-field can be found, for example, using the angular
spectrum representation of the CSD matrix~\cite{Tervo02}. We thus have, at any plane $z>0$,
\begin{align}
\boldmatrix{W}(\mathbf{r}_1,\mathbf{r}_2) &=\frac{1}{(2\pi)^4}
\iiiint_{-\infty}^\infty
\boldmatrix{A}(\boldsymbol{\kappa}_1,\boldsymbol{\kappa}_2)\nonumber\\
&\quad\times \exp\left[{\rm
i}(\mathbf{k}_2\cdot\mathbf{r}_2-\mathbf{k}_1^*\cdot\mathbf{r}_1)\right]
\mathrm{d}^2\kappa_1\,\mathrm{d}^2\kappa_2, \label{angspecdef}
\end{align}
where the angular correlation matrix (ACM)
\begin{align}
\boldmatrix{A}(\boldsymbol{\kappa}_1,\boldsymbol{\kappa}_2)
&=\iiiint_{-\infty}^\infty
\boldmatrix{W}(\boldsymbol{\rho}_1,\boldsymbol{\rho}_2,0)\nonumber\\
&\quad\times \exp\left[{\rm
i}(\boldsymbol{\kappa}_1\cdot\boldsymbol{\rho}_1-\boldsymbol{\kappa}_2\cdot\boldsymbol{\rho}_2)\right]
\mathrm{d}^2\rho_1\,\mathrm{d}^2\rho_2 \label{defangspec}
\end{align}
describes the correlations between vectorial plane-wave components. In the far zone~\cite{Tervo02}
\begin{align}
\boldmatrix{W}^\infty(r_1\mathbf{\hat{s}}_1,r_2\mathbf{\hat{s}}_2)
&=(2\pi k)^2\cos\theta_1\cos\theta_2\boldmatrix{A}(k\boldsymbol{\sigma}_1, k\boldsymbol{\sigma}_2) \nonumber\\
&\quad\times \frac{\exp\left[{\rm i}k(r_2-r_1)\right]}{r_1r_2}
\label{farfield}
\end{align}
and the Poynting vector takes the form~\cite{Tervo02}
\begin{align}
\mathbf{P}^\infty(r\mathbf{\hat{s}})
&=\frac{n\mathbf{\hat{s}}}{2}\sqrt{\frac{\epsilon_0}{\mu_0}}
\mathrm{tr}\,\boldmatrix{W}^\infty(r\mathbf{\hat{s}},r\mathbf{\hat{s}})\nonumber\\
&= \mathbf{\hat{s}}\cos^2\theta\frac{2n\pi^2 k^2}{r^2}
\sqrt{\frac{\epsilon_0}{\mu_0}}\,\mathrm{tr}\,\boldmatrix{A}(k\boldsymbol{\sigma},
k\boldsymbol{\sigma}), \label{Poynting}
\end{align}
where $n$ is the refractive index of the material, and $\epsilon_0$ and
$\mu_0$ are the vacuum permittivity and permeability, respectively.
Furthermore, the radiant intensity is
\begin{align}
J(r\mathbf{\hat{s}})&=\lim_{r\to\infty}[r^2\|\mathbf{P}^\infty(r\mathbf{\hat{s}})\|]\nonumber\\
&= 2n \pi^2 k^2   \cos^2\theta
\sqrt{\frac{\epsilon_0}{\mu_0}}\,\mathrm{tr}\,\boldmatrix{A}(k\boldsymbol{\sigma},
k\boldsymbol{\sigma}). \label{RadiantIntensity}
\end{align}

Owing to Eqs.~\eqref{Wpqdef} and \eqref{defangspec}, also ACM has a representation as a correlation
matrix:
\begin{align}
\boldmatrix{A}(\boldsymbol{\kappa}_1, \boldsymbol{\kappa}_2) =
\langle\mathbf{A}^\ast(\boldsymbol{\kappa}_1)
\mathbf{A}^\mathrm{T}(\boldsymbol{\kappa}_2)\rangle,
\label{angspeccorr}
\end{align}
where the components of $\mathbf{A}(\boldsymbol{\kappa})$ represent, componentwise, the angular
spectra~\cite{Mandel} of the electric-field realizations. It follows
from Eq.~\eqref{angspeccorr} that
$\boldmatrix{A}(\boldsymbol{\kappa}, \boldsymbol{\kappa})$ is
Hermitian, satisfying
\begin{align}
\boldmatrix{A}^\dagger(\boldsymbol{\kappa}_1, \boldsymbol{\kappa}_2)
=\boldmatrix{A}(\boldsymbol{\kappa}_2, \boldsymbol{\kappa}_1),
\label{A:Hermitian}
\end{align}
where the dagger denotes the adjoint matrix. It is also non-negative
definite in the sense that
\begin{align}
\iiiint\mathbf{a}^\dagger(\boldsymbol{\kappa}_1)
\boldmatrix{A}(\boldsymbol{\kappa}_1, \boldsymbol{\kappa}_2)
\mathbf{a}(\boldsymbol{\kappa}_2) \mathrm{d}^2\kappa_1
\mathrm{d}^2\kappa_2\ge0, \label{A:Nonneg}
\end{align}
where $\mathbf{a}(\boldsymbol{\kappa})$ is an arbitrary,
sufficiently well-behaved vector function of the same size as the
electric-field vector.

The following argument is essential for the conclusions of this paper:
the field in the far zone is well known to be a modulated outgoing spherical wave.
Thus, in spherical polar coordinates, the angular-spectrum vector is
two-dimensional, i.e.,
$\hat{\mathbf{s}}\cdot\mathbf{A}(k\boldsymbol{\sigma})=0$ for
every $\hat{\mathbf{s}}$. Hence $\boldmatrix{A}(\boldsymbol{\kappa}_1, \boldsymbol{\kappa}_2)$
is expressible as a $2\times 2$ matrix in these coordinates.

Since $\boldmatrix{A}(\boldsymbol{\kappa}_1, \boldsymbol{\kappa}_2)$
is a Hermitian, non-negative definite $2\times 2$
matrix, it has two non-negative real-valued eigenfunctions.
In view of Eqs.~\eqref{A:Hermitian} and~\eqref{A:Nonneg}, we have at $\boldsymbol{\kappa}_1=\boldsymbol{\kappa}_2=\boldsymbol{\kappa}$
the decomposition
\begin{align}
\boldmatrix{A}(\boldsymbol{\kappa}, \boldsymbol{\kappa})
&=\sum_{j=1}^2I_j(\boldsymbol{\kappa})\mathbf{F}_j^*
(\boldsymbol{\kappa})\mathbf{F}_j^\mathrm{T}(\boldsymbol{\kappa}),
\label{decomposition}
\end{align}
where $I_j(\boldsymbol{\kappa})$ are
the eigenvalues and $\mathbf{F}_f(\boldsymbol{\kappa})$ are the eigenvectors of
$\boldmatrix{A}(\boldsymbol{\kappa}, \boldsymbol{\kappa})$. The
eigenvectors may be assumed orthonormal, i.e.,
\begin{align}
\mathbf{F}_p^\dagger(\boldsymbol{\kappa})\mathbf{F}_q(\boldsymbol{\kappa})=\delta_{pq}.
\label{Orthogonality}
\end{align}
In other words, the matrix $\boldmatrix{A}(\boldsymbol{\kappa}, \boldsymbol{\kappa})$ can be
diagonalized. Evaluation of the trace in Eq.~\eqref{RadiantIntensity}
yields the explicit form for the radiant intensity:
\begin{equation}
J(r\mathbf{\hat{s}})=J_0 \cos^2\theta\left[I_1(k\boldsymbol{\sigma})
+I_2(k\boldsymbol{\sigma})\right],\label{RadiantModes}
\end{equation}
where $J_0 = 2n \pi^2 k^2\sqrt{\epsilon_0/\mu_0}$. Thus
it is proportional to the sum of the eigenvalues of the ACM at
$\boldsymbol{\kappa}_1=\boldsymbol{\kappa}_2$.

The decomposition in Eq.~\eqref{decomposition} is, in fact, valid
regardless of the chosen coordinate system; the eigenvalues remain invariant
in all unitary transformations, including simple coordinate transformations
between, e.g., Cartesian and spherical polar coordinate systems. As explicitly expressed in
Eqs.~\eqref{decomposition}, the polarization decomposition of ACM is generally direction-dependent:
the eigenvalues and/or the eigenvectors depend on $\hat{\mathbf{s}}$.

The physical meaning of Eq.~\eqref{decomposition} is clear:
In each direction in the far-zone, we may decompose the single-point
ACM into two mutually uncorrelated, orthogonal
polarization components. Moreover, owing to the factorized form
$I_j(\boldsymbol{\kappa})\mathbf{F}_j^*(\boldsymbol{\kappa})\mathbf{F}_j^\mathrm{T}(\boldsymbol{\kappa})$
of these components, they both represent fully polarized
fields. This can be verified from the well-known formula for
the (space--frequency domain) degree of polarization:
\begin{align}
P(\mathbf{r})=\left\{1-\frac{4\,\mathrm{det}\,\boldmatrix{W}(\mathbf{r},\mathbf{r})}%
{[\mathrm{tr}\,\boldmatrix{W}(\mathbf{r},\mathbf{r})]^2}\right\}^{1/2}.
\end{align}
In the far zone we have, using Eq.~(\ref{farfield}),
\begin{align}
P(r\mathbf{\hat{s}})=\left\{1-\frac{4\,\mathrm{det}\,\boldmatrix{A}(k\boldsymbol{\sigma},k\boldsymbol{\sigma})}%
{[\mathrm{tr}\,\boldmatrix{A}(k\boldsymbol{\sigma},k\boldsymbol{\sigma})]^2}\right\}^{1/2}.
\label{farzoneD}
\end{align}
With the aid of Eqs.~(\ref{decomposition}) and (\ref{Orthogonality}),
we then obtain
\begin{align}
P(r\mathbf{\hat{s}})=\left|\frac{I_1(k\boldsymbol{\sigma})-I_2(k\boldsymbol{\sigma})}
{I_1(k\boldsymbol{\sigma})+I_2(k\boldsymbol{\sigma})}\right|.\label{Idegpol}
\end{align}
Thus $P(r\mathbf{\hat{s}})=1$ for each individual polarization mode,
i.e., if either $I_1(\boldsymbol{\kappa})=0$ or
$I_2(\boldsymbol{\kappa})=0$. It is worth stressing that the
polarization decomposition presented above is analogous with the
well-known decomposition of a partially polarized plane wave into
two polarization modes: in particular, Eq.~(\ref{Idegpol}) is
analogous with Eq.~(6.3--31) in Ref.~\cite{Mandel}. In our case,
however, the direction-dependent eigenvalues are those of a
partially polarized and partially coherent field in the far zone.

To be able to include partial coherence (in addition to partial polarization) in the analysis,
we now assume that also the two-point ACM can be expressed
in the form
\begin{align}
\boldmatrix{A}(\boldsymbol{\kappa}_1, \boldsymbol{\kappa}_2)
=\boldmatrix{A}_1(\boldsymbol{\kappa}_1, \boldsymbol{\kappa}_2)+
\boldmatrix{A}_2(\boldsymbol{\kappa}_1, \boldsymbol{\kappa}_2),
\label{assumption}
\end{align}
where $\boldmatrix{A}_j(\boldsymbol{\kappa}_1,
\boldsymbol{\kappa}_2)$ have diagonal values $\boldmatrix{A}_j(\boldsymbol{\kappa},
\boldsymbol{\kappa})
=I_j(\boldsymbol{\kappa})\mathbf{F}_j^*(\boldsymbol{\kappa})\mathbf{F}_j^\mathrm{T}
(\boldsymbol{\kappa})$. In other words,
$\boldmatrix{A}(\boldsymbol{\kappa}_1, \boldsymbol{\kappa}_2)$ is
assumed to be expressible as a sum of two mutually uncorrelated, but
fully coherent and polarized modes even if $\boldsymbol{\kappa}_1\neq
\boldsymbol{\kappa}_2$. We stress that, while Eq.~(\ref{assumption})
does not follow from Eq.~(\ref{decomposition}), it typically holds.
We do not dwell into a detailed discussion of this point here, but
note that it is a nontrivial task to find counterexamples.

We proceed to investigate the angular correlation properties of the
class of fields described by Eq.~\eqref{assumption}. Without loss of
generality, we may employ spherical polar coordinates in the far
field, in which case $\boldmatrix{A}_j(\boldsymbol{\kappa}_1,
\boldsymbol{\kappa}_2)$ is a $2\times 2$ matrix. Let us denote by
$\boldmatrix{U}(\boldsymbol{\kappa})$ a unitary matrix whose columns
are the eigenvectors $\mathbf{F}_j(\boldsymbol{\kappa})$ of
$\boldmatrix{A}(\boldsymbol{\kappa}, \boldsymbol{\kappa})$ and by
$\boldmatrix{D}(\boldsymbol{\kappa},\boldsymbol{\kappa})$ a diagonal
matrix with elements equal to the eigenvalues
$I_j(\boldsymbol{\kappa})$ of $\boldmatrix{A}(\boldsymbol{\kappa},
\boldsymbol{\kappa})$. Then, equivalently with
Eq.~\eqref{decomposition}, we may write
\begin{align}
\boldmatrix{A}(\boldsymbol{\kappa}, \boldsymbol{\kappa})
=\boldmatrix{U}^*(\boldsymbol{\kappa})
\boldmatrix{D}(\boldsymbol{\kappa},\boldsymbol{\kappa})
\boldmatrix{U}^\mathrm{T}(\boldsymbol{\kappa})
\end{align}
and consequently
\begin{align}
\boldmatrix{U}^\mathrm{T}(\boldsymbol{\kappa})\boldmatrix{A}(\boldsymbol{\kappa},
\boldsymbol{\kappa})\boldmatrix{U}^*(\boldsymbol{\kappa}) =
\boldmatrix{D}(\boldsymbol{\kappa},\boldsymbol{\kappa}).
\label{singular11}
\end{align}
In view of Eq.~(\ref{assumption}) and the associated discussion,
$\boldmatrix{A}_j(\boldsymbol{\kappa}_1, \boldsymbol{\kappa}_2)$
represents the angular correlation between two plane-wave components
whose polarization states are described by deterministic vectors
$\mathbf{F}_j(\boldsymbol{\kappa}_1)$ and
$\mathbf{F}_j(\boldsymbol{\kappa}_2)$. Thus we may write, in analogy
with Eq.~(\ref{decomposition}),
\begin{align}
\boldmatrix{A}(\boldsymbol{\kappa}_1, \boldsymbol{\kappa}_2)
&=\sum_{j=1}^2G_j(\boldsymbol{\kappa}_1,\boldsymbol{\kappa}_2)
\mathbf{F}_j^*(\boldsymbol{\kappa}_1)
\mathbf{F}_j^\mathrm{T}(\boldsymbol{\kappa}_2),\label{decomposition2}
\end{align}
where $G_j(\boldsymbol{\kappa}_1,\boldsymbol{\kappa}_2)$ are scalar
(angular correlation) functions. It is seen by direct calculation
that $\boldmatrix{A}(\boldsymbol{\kappa}_1, \boldsymbol{\kappa}_2)$
has a representation similar to
Eq.~(\ref{singular11}),
\begin{align}
\boldmatrix{U}^\mathrm{T}(\boldsymbol{\kappa}_1)\boldmatrix{A}(\boldsymbol{\kappa}_1,
\boldsymbol{\kappa}_2)\boldmatrix{U}^*(\boldsymbol{\kappa}_2) =
\boldmatrix{D}(\boldsymbol{\kappa}_1, \boldsymbol{\kappa}_2),
\end{align}
where $\boldmatrix{D}(\boldsymbol{\kappa}_1, \boldsymbol{\kappa}_2)$
is a diagonal matrix with elements
$G_j(\boldsymbol{\kappa}_1,\boldsymbol{\kappa}_2)$. Thus
$\mathbf{F}_j(\boldsymbol{\kappa})$ and $G_j(\boldsymbol{\kappa}_1,\boldsymbol{\kappa}_2)$
can be determined by singular value decomposition of
$\boldmatrix{A}(\boldsymbol{\kappa}_1,\boldsymbol{\kappa}_2)$.

It can be shown, e.g., using Eq.~\eqref{A:Nonneg}
that the singular values $G_j(\boldsymbol{\kappa}_1,\boldsymbol{\kappa}_2)$ satisfy
\begin{align}
|G_j(\boldsymbol{\kappa}_1,\boldsymbol{\kappa}_2)|^2\le
I_j(\boldsymbol{\kappa}_1)I_j(\boldsymbol{\kappa}_2).
\end{align}
As a result, we may define the normalized angular correlation
functions
\begin{align}
g_j(\boldsymbol{\kappa}_1,\boldsymbol{\kappa}_2)
=\frac{G_j(\boldsymbol{\kappa}_1,\boldsymbol{\kappa}_2)}%
{\left[I_j(\boldsymbol{\kappa}_1)I_j(\boldsymbol{\kappa}_2)\right]^{1/2}}
\label{normG}
\end{align}
satisfying the inequalities
$0\le\left|g_j(\boldsymbol{\kappa}_1,\boldsymbol{\kappa}_2)\right|
\le 1$. If the correlations in the far zone are of
(generalized) Schell-model form, i.e.,
$g_j(\boldsymbol{\kappa}_1,\boldsymbol{\kappa}_2)
=g_j(\Delta\boldsymbol{\kappa})$, it follows from Eqs.~(\ref{decomposition2}) and (\ref{normG}) that
\begin{align}
\boldmatrix{A}(\boldsymbol{\kappa}_1, \boldsymbol{\kappa}_2)
&=\sum_{j=1}^2 g_j(\Delta\boldsymbol{\kappa})
\mathbf{f}_j^*(\boldsymbol{\kappa}_1)
\mathbf{f}_j^\mathrm{T}(\boldsymbol{\kappa}_2)
\label{A:StartingPoint}
\end{align}
with
\begin{equation}
\mathbf{f}_j(\boldsymbol{\kappa})
=[I_j(\boldsymbol{\kappa})]^{1/2}\mathbf{F}_j (\boldsymbol{\kappa}).
\label{HF}
\end{equation}
This is the electromagnetic extension of Eq.~\eqref{Aschell}.

\section{Elementary electric-field modes}
\label{s:elementary}

It follows from Eq.~\eqref{assumption} and the linearity of
Eq.~\eqref{defangspec} that the CSD has
the decomposition
\begin{align}
\boldmatrix{W}(\mathbf{r}_1,\mathbf{r}_2)
=\boldmatrix{W}_1(\mathbf{r}_1,\mathbf{r}_2)
+\boldmatrix{W}_2(\mathbf{r}_1,\mathbf{r}_2). \label{FieldSource}
\end{align}
Let us define the inverse Fourier transforms $\mathbf{e}_j(\mathbf{r})$ and $p_j(\boldsymbol{\rho})$
of the functions $\mathbf{f}_j(\boldsymbol{\kappa})$ and $g_j(\Delta\boldsymbol{\kappa})$ in
analogy with Eqs.~\eqref{IFT-e} and \eqref{IFT-p}. With a procedure similar to that used in
derivation of Eq.~\eqref{Scalarrep}, we can express the two terms in Eq.~(\ref{FieldSource}) in the form
\begin{align}
\boldmatrix{W}_j(\mathbf{r}_1,\mathbf{r}_2) =\iint_{-\infty}^\infty
p_j(\boldsymbol{\rho}')
\mathbf{e}_j^*(\mathbf{r}_1-\boldsymbol{\rho}' )
\mathbf{e}_j^\mathrm{T}(\mathbf{r}_2-\boldsymbol{\rho}' )
\mathrm{d}^2\rho' \label{Elementary}
\end{align}
in the half-space $z\geq 0$. In particular, Eq.~(\ref{Elementary})
is valid at the source plane $z = 0$, where
$\mathbf{e}_j(\mathbf{r}) = \mathbf{e}_j(\boldsymbol{\rho},0)$. This
result is the electromagnetic extension of the scalar elementary-mode
decomposition in Eq.~\eqref{Scalarrep}.

Let us next examine some general properties of Eq.~(\ref{Elementary}).
Equations \eqref{Orthogonality} and \eqref{HF}
together with
\begin{equation}
\mathbf{f}_j(\boldsymbol{\kappa}) = \iint_{-\infty}^\infty
\mathbf{e}_j(\mathbf{r}) \exp\left(-{\rm i}
\mathbf{k}\cdot\mathbf{r}\right)\,\mathrm{d}^2\rho
\label{AngspecGinv}
\end{equation}
lead to
\begin{equation}
\iiiint_{-\infty}^\infty
\mathbf{e}_1^\dagger(\boldsymbol{\rho}-\boldsymbol{\rho}',0)\mathbf{e}_2(\boldsymbol{\rho},0)
\exp\left(-{\rm
i}\boldsymbol{\kappa}\cdot\boldsymbol{\rho}'\right)\,\mathrm{d}^2\rho\,\mathrm{d}^2\rho'=0.
\end{equation}
Because of the exponential factor in the integral, $\mathbf{e}_1(\boldsymbol{\rho},0)$ and
$\mathbf{e}_2(\boldsymbol{\rho},0)$ are generally not orthogonal in
a point\-wise sense. For many paraxial fields, though, the functions
$\mathbf{f}_1(\boldsymbol{\kappa})$ and
$\mathbf{f}_2(\boldsymbol{\kappa})$ are globally orthogonal, i.e.,
$\mathbf{f}_1^\dagger(\boldsymbol{\kappa}_1)\mathbf{f}_2(\boldsymbol{\kappa}_2)=0$
for all $\boldsymbol{\kappa}_1$ and $\boldsymbol{\kappa}_2$. In
this special case the point\-wise orthogonality of
$\mathbf{e}_1(\boldsymbol{\rho},0)$ and
$\mathbf{e}_2(\boldsymbol{\rho},0)$ follows from
Eqs.~\eqref{Orthogonality}, \eqref{HF}, and \eqref{AngspecGinv}.
This property holds also for significant classes of non-paraxial
fields, as will be demonstrated in Sections \ref{s:rotsym} and
\ref{s:cosine}. However, it is not essential for practical implementation of
the propagation algorithm developed in this paper.

Since the functions $\mathbf{e}_j(\mathbf{r})$ represent fully coherent fields, they
obey the Helmholtz equation
\begin{align}
\nabla^2\mathbf{e}_j(\mathbf{r}) +k^2\mathbf{e}_j(\mathbf{r})=0.
\label{HelmholtzG}
\end{align}
Furthermore, the CSD matrix obeys the divergence
equation~\cite{Tervo04}
\begin{align}
\nabla_1^\mathrm{T}\boldmatrix{W}(\mathbf{r}_1,\mathbf{r}_2)=\mathbf{0},
\label{DivergenceW}
\end{align}
where the subscript 1 denotes differentiation with respect to
$\mathbf{r}_1$. Together with Eqs.~\eqref{FieldSource} and
\eqref{Elementary}, this implies that
\begin{align}
\nabla\cdot\mathbf{e}_j(\mathbf{r})=0. \label{DivergenceG}
\end{align}
In view of Eqs.~\eqref{HelmholtzG} and \eqref{DivergenceG}, the
functions $\mathbf{f}_j(\mathbf{r})$ behave exactly as the electric
field vector in the space--frequency domain. Hence the
matrix-functions
\begin{align}
\boldmatrix{W}_j^{\rm e}(\mathbf{r}_1,\mathbf{r}_2)
=\mathbf{e}_j^*(\mathbf{r}_1)\mathbf{e}_j^\mathrm{T}(\mathbf{r}_2)
\label{elfact}
\end{align}
may be called the \emph{elementary electric-field modes} of the
field. Using these modes, we can write Eqs.~\eqref{FieldSource} and
\eqref{Elementary} in a more compact form
\begin{equation}
\boldmatrix{W}(\mathbf{r}_1,\mathbf{r}_2)=\sum_{j=1}^2\iint_{-\infty}^\infty
p_j(\boldsymbol{\rho}')\boldmatrix{W}_j^{\rm
e}(\mathbf{r}_1-\boldsymbol{\rho}',\mathbf{r}_2-\boldsymbol{\rho}')
\mathrm{d}^2\rho' \label{ElementaryB}
\end{equation}
and express the spectral density of the field as
\begin{equation}
S(\mathbf{r})=\sum_{j=1}^2\iint_{-\infty}^\infty
p_j(\boldsymbol{\rho}')S_j^{\rm e} (\mathbf{r}-\boldsymbol{\rho}')
\mathrm{d}^2\rho', \label{SDensity}
\end{equation}
where $S_j^{\rm e}(\mathbf{r})=\mathrm{tr}\,\boldmatrix{W}_j^{\rm
e}(\mathbf{r},\mathbf{r})$ are the spectral densities of the two
elementary electric-field modes.

In analogy with scalar theory~\cite{Vahimaa}, the field in
Eq.~\eqref{ElementaryB} is understood to consist of a weighted
continuum of identical, laterally shifted elementary modes. The main
difference between the scalar and electromagnetic descriptions is
that the electromagnetic field consists of two (uncorrelated) sets
of elementary modes, whose polarization states in the far-zone are
orthogonal and whose radiant-intensity distributions are in general
different. Owing to the factorized form of $\boldmatrix{W}_j^{\rm
e}(\mathbf{r}_1,\mathbf{r}_2)$ in Eq.~\eqref{elfact}, each
elementary mode is completely coherent~\cite{Setala2004} in the
sense of the space--frequency analog of the degree of coherence for
electromagnetic fields put forward in Ref.~\cite{Tervo2003}:
\begin{align}
\mu(\mathbf{r}_1,\mathbf{r}_2)
=\{\mathrm{tr}\left[\boldsymbol{\mu}(\mathbf{r}_1,\mathbf{r}_2)%
\boldsymbol{\mu}(\mathbf{r}_2,\mathbf{r}_1)\right]\}^{1/2}.
\label{scalarmu}
\end{align}
Here
\begin{align}
\boldsymbol{\mu}(\mathbf{r}_1,\mathbf{r}_2)=
\frac{\boldmatrix{W}(\mathbf{r}_1,\mathbf{r}_2)}%
{\left[S(\mathbf{r}_1)S(\mathbf{r}_2)\right]^{1/2}} \label{boldmu}
\end{align}
is the normalized CSD matrix. In general, all
$3\times 3$ matrix elements of $\boldmatrix{W}_j^{\rm
e}(\mathbf{r}_1,\mathbf{r}_2)$ are non-zero and spatially varying.

Analogously with Eqs.~\eqref{scalarmu} and \eqref{boldmu}, we can
define the degree of angular coherence
\begin{equation}
\alpha(\boldsymbol{\kappa}_1,\boldsymbol{\kappa}_2)
=\{\mathrm{tr}\left[\boldsymbol{\alpha}(\boldsymbol{\kappa}_1,\boldsymbol{\kappa}_2)
\boldsymbol{\alpha}(\boldsymbol{\kappa}_2,\boldsymbol{\kappa}_1)\right]\}^{1/2}
\label{scalaralpha}
\end{equation}
where
\begin{equation}
\boldsymbol{\alpha}(\boldsymbol{\kappa}_1,\boldsymbol{\kappa}_2)=
\frac{\boldmatrix{A}(\boldsymbol{\kappa}_1,\boldsymbol{\kappa}_2)}
{\left[{\rm tr}\,\boldmatrix{A}(\boldsymbol{\kappa}_1,\boldsymbol{\kappa}_1)
{\rm tr}\,\boldmatrix{A}(\boldsymbol{\kappa}_2,\boldsymbol{\kappa}_2)\right]^{1/2}}
\label{boldalpha}
\end{equation}
is normalized angular correlation matrix.

To conclude this section we provide a convenient, fully general
series representation for the electric-field modes. In the far zone,
where the field is transverse to the local propagation direction and
thus fluctuates locally in the plane defined by the spherical polar
unit vectors $\boldsymbol{\hat{\theta}}$ and
$\boldsymbol{\hat{\psi}}$, it is natural to express the basis
vectors in the form
\begin{equation}
\mathbf{f}_j(\boldsymbol{\kappa})=
f_{j,\theta}(\boldsymbol{\kappa})\boldsymbol{\hat{\theta}}+
f_{j,\psi}(\boldsymbol{\kappa})\boldsymbol{\hat{\psi}}.
\label{GeneralSphericalAzimuthal}
\end{equation}
Using spherical
polar coordinates $\left(\theta,\psi\right)$ for the spatial
frequencies and circular cylindrical coordinates
$\left(\rho,\phi,z\right)$ for the position vector (see
Fig.~\ref{f:notations}), we then have (after somewhat lengthy
calculations outlined in Appendix \ref{A1})
\begin{align}
\mathbf{e}_j(\mathbf{r})
&=\frac{k}{2\pi }\sum_{m=-\infty}^\infty {\rm i}^m \exp({\rm
i}m\phi)\int_0^{\pi/2}
\exp\left({\rm i} kz\cos\theta\right)\nonumber\\
&\quad\times \biggl\{ \boldsymbol{\hat{\varrho}} \left[ -\frac{
m}{\rho}f_{j,\psi,m}(\theta) -{\rm i} \cos\theta
f_{j,\theta,m}(\theta) \frac{\mathrm{d}}{\mathrm{d\rho}}
\right]\nonumber\\
&\qquad+ \boldsymbol{\hat{\phi}} \left[ \frac{ m}{\rho}\cos\theta
f_{j,\theta,m}(\theta) -{\rm i} f_{j,\psi,m}(\theta)
\frac{\mathrm{d}}{\mathrm{d\rho}}
\right]\nonumber\\
&\qquad- \mathbf{\hat{z}} k \sin^2\theta f_{j,\theta,m}(\theta)
\biggr\}J_m(k\rho\sin\theta)\,\mathrm{d}\theta, \label{Cylinder}
\end{align}
where
\begin{equation}
f_{j,\xi,m}(\theta) =\frac{1}{2\pi}\int_0^{2\pi}
f_{j,\xi}(\theta,\psi)\exp\left(-{\rm i}m\psi\right)\,\mathrm{d}\psi
\label{fouriercoeff}
\end{equation}
are the azimuthal Fourier coefficients of $f_{j,\xi}(\theta,\psi)$
and $\xi$ stands for either $\theta$ or $\psi$. Note that the upper
limit of the integral in Eq.~\eqref{Cylinder} is set to $\pi/2$.
Thus the elementary electric-field modes are taken to contain only
propagating waves, i.e., information that can be gathered from
far-zone measurements. As a result, the propagation method
considered here is not suitable for modeling near-field phenomena
(fields at distances of the order of one wavelength from the plane
$z=0$).

\section{Rotationally symmetric fields}
\label{s:rotsym}

Let us assume that the CSD at $z = 0$ is rotationally symmetric about the $z$ axis.
Then also the ACM is rotationally symmetric
and the polarization basis vectors
$\mathbf{f}_1(\boldsymbol{\kappa})$ and
$\mathbf{f}_2(\boldsymbol{\kappa})$ are rotationally invariant,
i.e., their $\theta$ and $\psi$ components
$f_{j,\theta}(\theta,\psi)$ and $f_{j,\psi}(\theta,\psi)$ are
independent on the azimuthal angle $\psi$. Hence only the
zeroth-order Fourier coefficients $f_{j,\xi,0}(\theta)$ in
Eq.~\eqref{fouriercoeff} are non-zero and it follows from
Eqs.~\eqref{Cylinder}
 and \eqref{BesselDiff} that
\begin{align}
\mathbf{e}_j(\mathbf{r})
&=\frac{k^2}{2\pi}\int_0^{\pi/2}
\sin\theta\exp\left({\rm i} kz\cos\theta \right)\nonumber\\
&\quad\times \biggl\{ {\rm i}J_1(k\rho\sin\theta) \left[ \cos\theta
f_{j,\theta,0}(\theta)\boldsymbol{\hat{\rho}}
+f_{j,\psi,0}(\theta)\boldsymbol{\hat{\phi}}
\right]\nonumber\\
&\qquad- \sin\theta
J_0(k\rho\sin\theta)f_{j,\theta,0}(\theta)\mathbf{\hat{z}} \biggr\}
\,\mathrm{d}\theta.
\end{align}
Thus, as expected, the elementary electric-field modes are also
rotationally symmetric about the $z$ axis. In particular, if the
basis vectors are parallel to $\boldsymbol{\hat{\theta}}$ and
$\boldsymbol{\hat{\psi}}$,
\begin{subequations}
\begin{align}
\mathbf{e}_1(\mathbf{r})
&=\frac{k^2}{2\pi}\int_0^{\pi/2}f_{j,\theta,0}(\theta)
\sin\theta \nonumber\\
&\quad\times \left[ {\rm
i}\boldsymbol{\hat{\rho}}J_1(k\rho\sin\theta) \cos\theta -
\mathbf{\hat{z}}\sin\theta J_0(k\rho\sin\theta) \right]
\nonumber\\
&\quad\times\exp\left({\rm i} kz\cos\theta \right)\,\mathrm{d}\theta,\\
\mathbf{e}_2(\mathbf{r}) &=\boldsymbol{\hat{\phi}}\frac{{\rm
i}k^2}{2\pi}\int_0^{\pi/2}f_{j,\psi,0}(\theta) \sin\theta
J_1(k\rho\sin\theta)
\nonumber\\
&\quad\times\exp\left({\rm i} kz\cos\theta
\right)\,\mathrm{d}\theta.
\end{align}
\label{GeneralRadialAzimuthal}
\end{subequations}
The elementary modes $\mathbf{e}_1(\mathbf{r})$ and
$\mathbf{e}_2(\mathbf{r})$ are now radially and azi\-muthally
polarized fields, respectively. Hence they are point\-wise
orthogonal, regardless of whether the field is paraxial or not.
Since these modes have no azimuthal phase variation, they possess no
phase singularity (vortex) at $\rho=0$. Thus, even if the axial
field vanished $z=0$ (as turns out to be often the case for some of
the field components), such a zero does not propagate.

\section{Quasi-homogeneous sources}
\label{s:quasihom}

Assume next that the variations of the spectral density
$S(\boldsymbol{\rho},0)$ at the source plane are slow compared to
the variations of the degree of coherence
$\mu(\boldsymbol{\rho}_1,0,\boldsymbol{\rho}_2,0)$. Moreover, let
the correlations at $z = 0$ be of the
Schell-model form, i.e., depend only on
$\Delta\boldsymbol{\rho}=\boldsymbol{\rho}_2-\boldsymbol{\rho}_1$.
Such a planar source is said to be quasi-homogeneous~\cite{Mandel}. We may then approximate
$S(\boldsymbol{\rho}_1,0)\approx S(\boldsymbol{\rho}_2,0)\approx
S(\boldsymbol{\bar{\rho}},0)$, where
$\boldsymbol{\bar{\rho}}=(\boldsymbol{\rho}_1+\boldsymbol{\rho}_2)/2$,
and write
\begin{align}
\boldmatrix{W}(\boldsymbol{\rho}_1,\boldsymbol{\rho}_2,0)\approx
S(\boldsymbol{\bar{\rho}},0)
\boldsymbol{\mu}(\Delta\boldsymbol{\rho},0). \label{quasi}
\end{align}
Inserting Eq.~\eqref{quasi} into Eq.~\eqref{defangspec} and defining $\boldsymbol{\bar{\kappa}}=(\boldsymbol{\kappa}_1
+\boldsymbol{\kappa}_2)/2$ yields
\begin{align}
\boldmatrix{A}(\boldsymbol{\kappa}_1,\boldsymbol{\kappa}_2)
=\tilde{S}(\Delta\boldsymbol{\kappa})
\tilde{\boldsymbol{\mu}}(\boldsymbol{\bar{\kappa}}),
\label{AngSpecQuasi}
\end{align}
where
\begin{equation}
\tilde{S}(\Delta\boldsymbol{\kappa}) = \iint_{-\infty}^\infty
S(\boldsymbol{\bar\rho},0)\exp\left(-{\rm
i}\Delta\boldsymbol{\kappa}\cdot\boldsymbol{\bar\rho}\right){\mathrm
d}^2\bar\rho,
\end{equation}
and
\begin{equation}
\tilde{\boldsymbol{\mu}}(\boldsymbol{\bar\kappa}) =
\iint_{-\infty}^\infty
\boldsymbol{\mu}(\Delta\boldsymbol{\rho},0)\exp\left(-{\rm
i}\boldsymbol{\bar\kappa}\cdot\Delta\boldsymbol{\rho}\right){\mathrm
d}^2\Delta\rho.
\end{equation}
Since
\begin{eqnarray}
\lefteqn{\mathrm{tr}\,\boldmatrix{A}(\boldsymbol{\kappa},\boldsymbol{\kappa})=
\tilde{S}(\boldsymbol{0})\,\mathbf{tr}\,\tilde{\boldsymbol{\mu}}(\boldsymbol{\kappa})}\nonumber\\
& & = \tilde{S}(\boldsymbol{0})\iint_{-\infty}^\infty
\mathrm{tr}\,\boldsymbol{\mu}(\Delta\boldsymbol{\rho},0)\exp\left(-{\rm
i}\boldsymbol{\kappa}\cdot\Delta\boldsymbol{\rho}\right){\mathrm
d}^2\Delta\rho,\nonumber\\
\end{eqnarray}
it follows from Eq.~(\ref{RadiantIntensity}) that the
radiant intensity produced by a quasi\-homogeneous electromagnetic
source depends only on the correlation properties of the source
field, in complete analogy with the scalar case~\cite{Mandel}.

Inserting from Eq.~\eqref{quasi} into Eq.~\eqref{boldalpha} and recalling that $\tilde{\mu}$ is a wide function compared to $\tilde{S}$, we have
\begin{equation}
\boldsymbol{\alpha}({\boldsymbol{\kappa}_1,\boldsymbol{\kappa}_2) = \frac{\tilde{S}(\Delta\boldsymbol{\kappa})}{\tilde{S}(\boldsymbol{0})}\frac{\tilde{\mu} (\bar{\boldsymbol{\kappa}})}{\tilde{\mu} (\boldsymbol{0})}}.
\end{equation}
Using Eq.~\eqref{scalaralpha} and noting that $\tilde{S}$ is real because $S$ is non-negative, we obtain
\begin{equation}
\alpha(\boldsymbol{\kappa}_1,\boldsymbol{\kappa}_2) = \frac{\tilde{S}(\Delta\boldsymbol{\kappa})}{\tilde{S}(\boldsymbol{0})}\frac{\left\{{\rm tr}\left[\tilde{\mu} (\bar{\boldsymbol{\kappa}})\right]^2\right\}^{1/2}}{{\rm tr}\,\tilde{\mu} (\bar{\boldsymbol{\kappa}})}.
\label{alpha-2}
\end{equation}
This expression can be cast into a more transparent form using the far-zone degree of polarization defined in Eq.~\eqref{farzoneD}, which can be written equivalently in the form
\begin{equation}
P(\boldsymbol{\kappa}) = \left\{\frac{2\, {\rm tr}\left[\tilde{\boldsymbol{\mu}}(\boldsymbol{\kappa})\right]^2}{
\left[{\rm tr}\,\tilde{\boldsymbol{\mu}}(\boldsymbol{\kappa})\right]^2}-1\right\}^{1/2}.
\end{equation}
We then have, from Eq.~\eqref{alpha-2},
\begin{equation}
\alpha(\boldsymbol{\kappa}_1,\boldsymbol{\kappa}_2) = \frac{\tilde{S}(\Delta\boldsymbol{\kappa})}{\tilde{S}(\boldsymbol{0})}
\left[\frac{D^2(\bar{\boldsymbol{\kappa}}) +1}{2}\right]^{1/2}.
\label{qhffD}
\end{equation}
The first fraction in this expression is equal to the scalar degree of angular coherence. The second fraction, however, is a polarization-dependent modulating term that depends on the source-plane correlations.

The assumption that the source is quasi\-homogeneous simplifies
decisively the elementary-mode decomposition of the field in the
scalar case~\cite{Vahimaa}, and the same is true in the
electromagnetic case. It follows from Eqs.~\eqref{decomposition}, \eqref{HF},
and~\eqref{AngspecGinv} that
\begin{align}
&\sum_{j=1}^2\iint_{-\infty}^\infty
\mathbf{e}_j^*(\boldsymbol{\rho}_1-\boldsymbol{\rho}',0)
\mathbf{e}_j^\mathrm{T}(\boldsymbol{\rho}_2-\boldsymbol{\rho}',0)\,
\mathrm{d}^2\rho'\nonumber\\
&=\frac{1}{(2\pi)^2}\iint_{-\infty}^\infty
\boldmatrix{A}(\boldsymbol{\kappa}, \boldsymbol{\kappa})
\exp(i\boldsymbol{\kappa}\cdot\Delta\boldsymbol{\rho})\,
\mathrm{d}^2\kappa\nonumber\\
&=\tilde{S}(\mathbf{0})
\boldsymbol{\mu}(\Delta\boldsymbol{\rho},0),\label{quasirelation}
\end{align}
where, in the last step, we  have used Eq.~\eqref{AngSpecQuasi}.
Comparing Eqs.~\eqref{quasi} and~\eqref{quasirelation} yields
\begin{align}
&\boldmatrix{W}(\boldsymbol{\rho}_1,\boldsymbol{\rho}_2,0)\approx
\frac{S(\boldsymbol{\bar{\rho}},0)}%
{\tilde{S}(\mathbf{0})}\nonumber\\
&\times\sum_{j=1}^2\iint_{-\infty}^\infty
\mathbf{e}_j^*(\boldsymbol{\rho}_1-\boldsymbol{\rho}',0)
\mathbf{e}_j^\mathrm{T}(\boldsymbol{\rho}_2-\boldsymbol{\rho}',0)\,
\mathrm{d}^2\rho'.
\end{align}
Thus the weight function no longer appears inside the integral and,
irrespective of the spatial distribution of the spectral density at
the source plane, the field characteristics can be determined the
propagating a convolution integral involving the elementary modes
only.

\section{Sources with cosine-power radiant intensity}
\label{s:cosine}

Let us consider the rotationally symmetric case with radially
and azimuthally polarized basis vectors $\mathbf{F}_1(\theta,\psi) = \boldsymbol{\hat{\theta}}$,
$\mathbf{F}_2(\theta,\psi) = \boldsymbol{\hat{\psi}}$, and corresponding eigenvalues
$I_1(\theta,\psi) = A_1^2\cos^{a-2}\theta$, $I_1(\theta,\psi) = A_2^2\cos^{b-2}\theta$,
where $A_1$ and $A_2$ are arbitrary (real) functions of frequency. Then, in view of Eq.~\eqref{RadiantModes},
the radiant intensity is a superposition of two $\cos^n\theta$ type contributions, one radially and
the other azimuthally polarized:
\begin{equation}
J(\theta,\psi) = J_0\left[A_1^2\cos^a\theta+A_2^2\cos^b\theta\right].
\end{equation}
The elementary electric-field modes in the far zone are, according to Eq.~\eqref{HF},
\begin{subequations}
\begin{align}
\mathbf{f}_1(\theta,\psi) & = \boldsymbol{\hat{\theta}}A_1\cos^{a/2-1}\theta,\\
\mathbf{f}_2(\theta,\psi) & = \boldsymbol{\hat{\psi}}A_2\cos^{b/2-1}\theta.
\end{align}
\end{subequations}
Using Eqs.~\eqref{GeneralSphericalAzimuthal} and \eqref{fouriercoeff} we see that the non-vanishing
Fourier coefficients are $f_{1,\theta,0} = A_1\cos^{a/2-1}\theta$ and
$f_{2,\psi,0} = A_2\cos^{b/2-1}\theta$. Inserting these into Eqs.~\eqref{GeneralRadialAzimuthal}
and applying~\eqref{IntegralSolution2} derived in Appendix \ref{A2}, we obtain the source-plane
elementary field modes in the form
\begin{subequations}
\begin{align}
\mathbf{e}_1(\rho,0) &=A_1\frac{{\rm i}k^2}{8\sqrt{\pi}}\left[
\boldsymbol{\hat{\rho}}\frac{k\rho}{2}\Gamma\left(\frac{1}{2}+\frac{a}{4}\right)\right.\nonumber\\
&\quad\times
\sideset{_1}{_2}\ER\left(\frac{3}{2};2,2+\frac{a}{4};-\frac{k^2\rho^2}{4}\right)\nonumber\\
&\quad-\left.\mathbf{\hat{z}}\Gamma\left(\frac{a}{4}\right)
\sideset{_1}{_2}\ER\left(\frac{3}{2};1,\frac{3}{2}+\frac{a}{4};-\frac{k^2\rho^2}{4}\right)\right],\\
\mathbf{e}_2(\rho,0) &=\boldsymbol{\hat{\phi}}A_2\frac{{\rm
i}k^3\rho}{16\sqrt{\pi}}
\Gamma\left(\frac{b}{4}\right)\nonumber\\
&\quad\times
\sideset{_1}{_2}\ER\left(\frac{3}{2};2,\frac{3}{2}+\frac{b}{4};-\frac{k^2\rho^2}{4}\right),
\end{align}
\label{spcosmodes}
\end{subequations}
where $\Gamma$ is the Gamma function and $\sideset{_1}{_2}\ER$ is the
regularized hypergeometric function (see Appendix \ref{A2}).

Figures~\ref{Cos1}--\ref{Cos3} illustrate the radial dependence of
the elementary-field components and the function
\begin{equation}
w(\rho,0)=\left[\|\mathbf{e}_1(\rho,0)\|^2+\|\mathbf{e}_2(\rho,0)\|^2\right]^{1/2}
\end{equation}
for different values of $a = b = n$, with $A_1=A_2=-{\rm i}k^{-2}$.

\begin{figure}[!h]
\psfrag{r}{\hspace{-1mm}$k\rho$}\psfrag{e}{\hspace{-7mm}amplitude}
\psfrag{0}{$0$}\psfrag{5}{$5$}\psfrag{10}{$10$}\psfrag{15}{$15$}\psfrag{20}{$20$}
\psfrag{0.1}{\hspace{2mm}$0.1$}\psfrag{0.2}{\hspace{2mm}$0.2$}\psfrag{0.3}{\hspace{2mm}$0.3$}
\psfrag{-0.1}{\hspace{2mm}$-0.1$}\psfrag{-0.2}{\hspace{2mm}$-0.2$}\psfrag{-0.3}{\hspace{2mm}$-0.3$}
\centering
\includegraphics[width=\columnwidth]{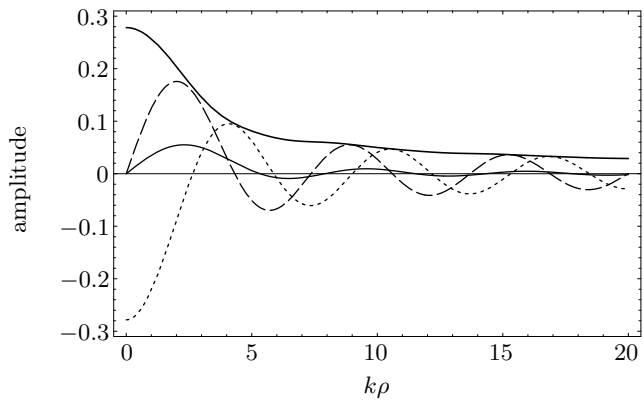}
\caption{Relative amplitudes of the radial (solid line), azimuthal
(dashed line), and longitudinal (dotted line) components of the
elementary fields as a function of the normalized radial coordinate
$k\rho$, as well as the function $w(k\rho)$ (thick solid line) for
$n=1$, which corresponds to a Lambertian source.} \label{Cos1}
\end{figure}

\begin{figure}[!h]
\psfrag{r}{\hspace{-1mm}$k\rho$}\psfrag{e}{\hspace{-7mm}amplitude}
\psfrag{0}{$0$}\psfrag{5}{$5$}\psfrag{10}{$10$}\psfrag{15}{$15$}\psfrag{20}{$20$}
\psfrag{0.1}{\hspace{2mm}$0.1$}\psfrag{0.05}{\hspace{2mm}$0.05$}
\psfrag{-0.1}{\hspace{2mm}$-0.1$}\psfrag{-0.05}{\hspace{2mm}$-0.05$}
\centering
\includegraphics[width=\columnwidth]{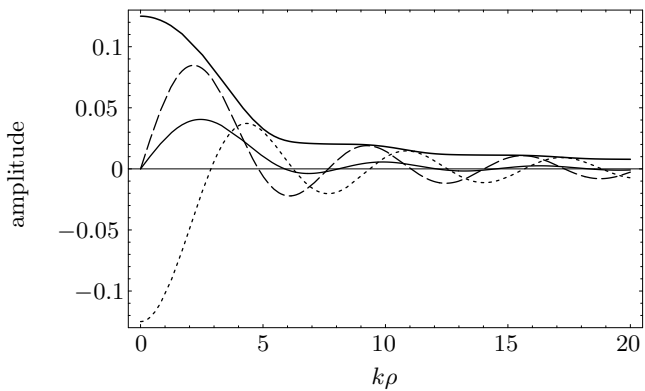}
\caption{Same as Fig.~\ref{Cos1}, but with $n=2$, which corresponds
to an incoherent source in scalar theory.} \label{Cos2}
\end{figure}

\begin{figure}[!h]
\psfrag{r}{\hspace{-1mm}$k\rho$}\psfrag{e}{\hspace{-7mm}amplitude}
\psfrag{0}{$0$}\psfrag{5}{$5$}\psfrag{10}{$10$}\psfrag{15}{$15$}\psfrag{20}{$20$}
\psfrag{0.02}{\hspace{2mm}$0.02$}\psfrag{0.04}{\hspace{2mm}$0.04$}
\psfrag{-0.02}{\hspace{2mm}$-0.02$}\psfrag{-0.04}{\hspace{2mm}$-0.04$}
\centering
\includegraphics[width=\columnwidth]{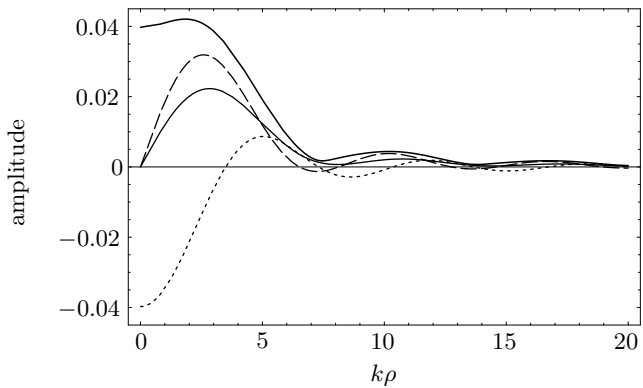}
\caption{Same as Fig.~\ref{Cos1}, but with $n=5$. Thus the source
has a somewhat directional radiation pattern.} \label{Cos3}
\end{figure}
\psfrag{0.0025}{\hspace{4mm}$0.0025$}\psfrag{0.005}{\hspace{3.5mm}$0.005$}
\psfrag{0.0075}{\hspace{4mm}$0.0075$}\psfrag{0.01}{\hspace{3mm}$0.01$}
\psfrag{-0.0025}{\hspace{4mm}$-0.0025$}\psfrag{-0.005}{\hspace{3.5mm}$-0.005$}

\begin{figure}[!h]
\psfrag{r}{\hspace{-1mm}$k\rho$}\psfrag{e}{\hspace{-7mm}amplitude}
\psfrag{0}{$0$}\psfrag{5}{$5$}\psfrag{10}{$10$}\psfrag{15}{$15$}\psfrag{20}{$20$}
\centering
\includegraphics[width=\columnwidth]{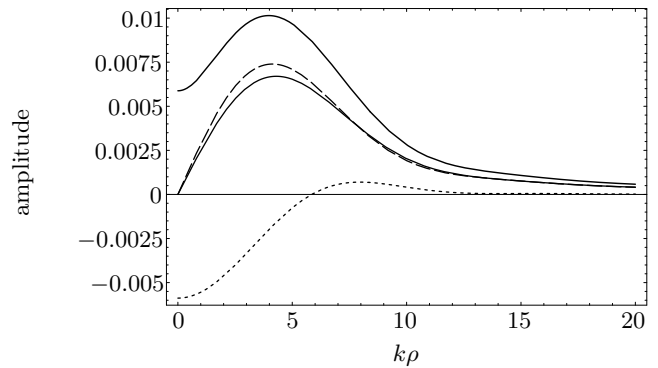}
\caption{Same as Fig.~\ref{Cos3}, but with $n=20$. The radiation
pattern is increasingly directional and could be produced
approximately by a LED with an integrated collimating lens.}
\label{Cos4}
\end{figure}

\section{Illustration: LED model}
\label{s:LED}

\begin{figure}[!h]
\psfrag{n}{\hspace{-1mm}$n_{\rm s}$}\psfrag{x}{$\theta$}
\psfrag{y}{\hspace{-7mm}$J(\theta)$, $P(\theta)$}
\psfrag{p}{pn}\psfrag{a}{(a)}\psfrag{b}{(b)}\psfrag{z}{$z$}
\psfrag{r}{$J_{\rm azi}(\theta)$}\psfrag{c}{$J_{\rm rad}(\theta)$}
\psfrag{0}{$0$}\psfrag{0.25}{$0.25$}\psfrag{0.5}{$0.5$}\psfrag{0.75}{$0.75$}\psfrag{1}{$1$}
\psfrag{1.25}{$1.25$}\psfrag{1.5}{$1.5$}\psfrag{0.2}{\hspace{2mm}$0.2$}\psfrag{0.4}{\hspace{2mm}$0.4$}
\psfrag{0.6}{\hspace{2mm}$0.6$}\psfrag{0.8}{\hspace{2mm}$0.8$}
\centering
\includegraphics[width=0.8\columnwidth]{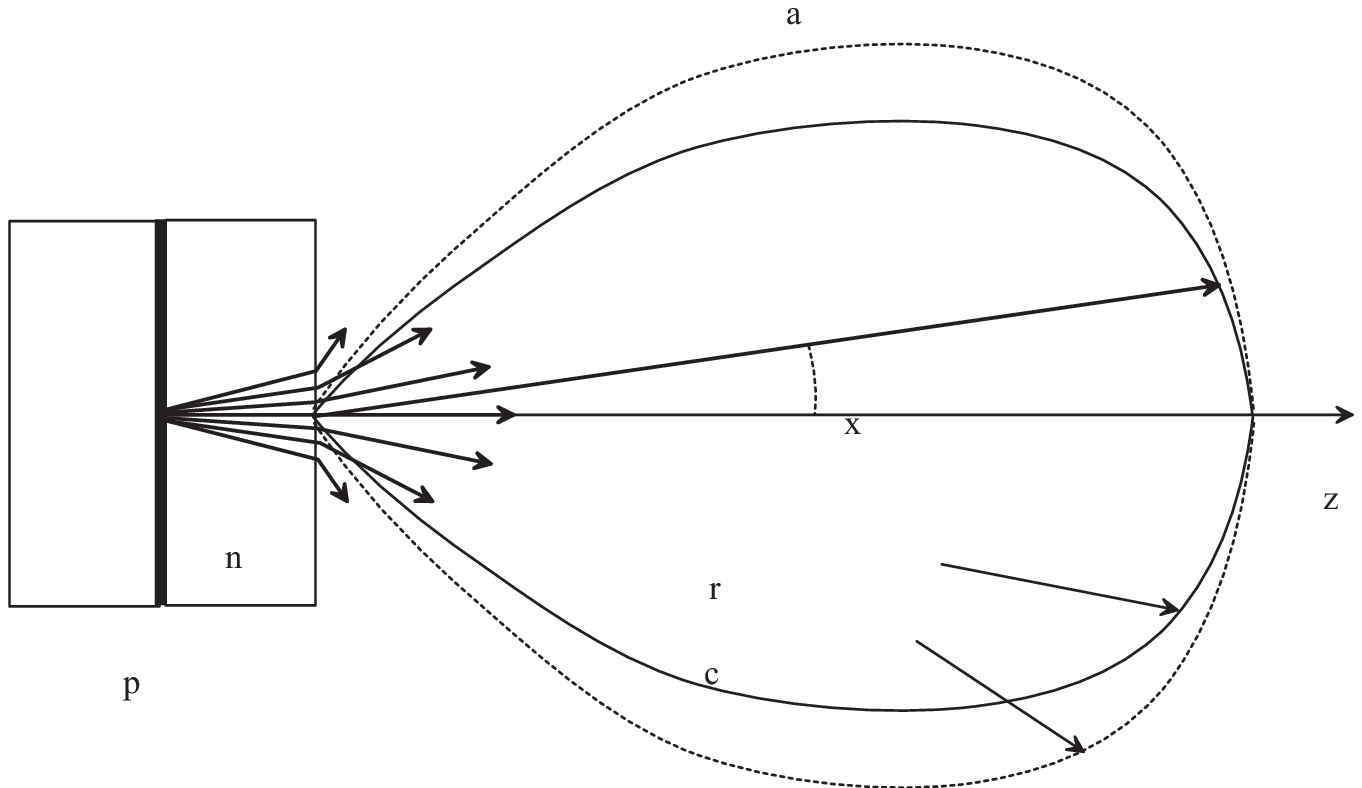}
\includegraphics[width=\columnwidth]{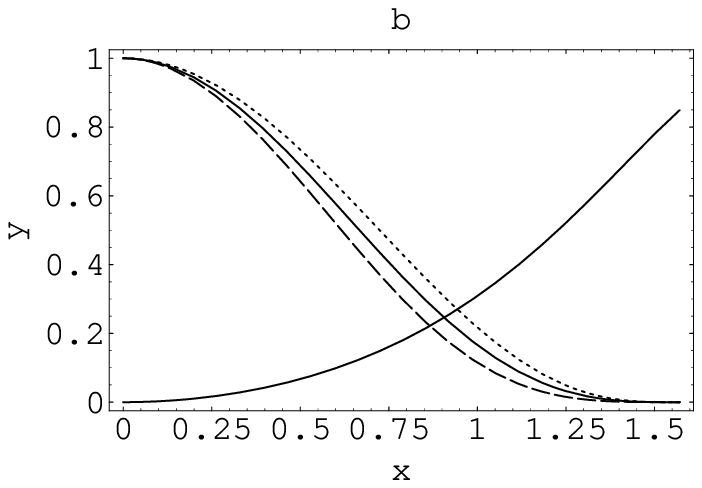}
\caption{(a) A generic geometry of a broad-area surface-emitting
LED: pn is the active emitting area and $n$ is the refractive index
of the semiconductor material. The solid and dashed curves
illustrate the azimuthally and radially polarized contributions to
the radiant intensity distribution $J(\theta)$.
(b)~Geometrical-optics predictions of the azimuthally (dashed curve)
and radially (dotted curve) polarized contributions to the radiant
intensity, their average (solid curve), and the degree of
polarization $P(\theta)$ in the far zone (thick solid curve).}
\label{f:LEDgeom}
\end{figure}

Let us consider a simple model for a rotationally symmetric surface-emitting LED
illustrated in Fig.~\ref{f:LEDgeom}a, where the primary light-emitting region is
planar (such as a quantum well) and buried inside a semiconductor
material of refractive index $n_{\rm s}$. Each primary source point is assumed to radiate
(independently) a spherical wave. If we denote the propagation angle inside the semiconductor material
by $\theta^\prime$, the radiant intensity may be expressed as a sum of radially
and azimuthally polarized contributions $J^{({\rm i})}_j(\theta^\prime)$, $j = 1,2$:
\begin{equation}
J^{({\rm i})}_j(\theta^\prime) = J_{0,j}^{({\rm i})}\cos^2\theta^\prime I_j^{({\rm i})}(\theta^\prime) = J_{0,j}^{({\rm i})}\cos^2\theta^\prime
\left|A_j^{({\rm i})}(\theta^\prime)\right|^2,
\end{equation}
where $A_j^{({\rm i})}(\theta^\prime)$ is the complex amplitude of the plane wave in direction $\theta^\prime$.
If the distance between the pn plane and the semiconductor-air interface is large compared to
$\lambda$, the local plane wave approximation is valid at the semiconductor-air interface.
Thus the output radiant intensity takes the form
\begin{equation}
J_j(\theta) = J_{0,j}\cos^2\theta I_j(\theta) = J_{0,j}\cos^2\theta
\left|A_j(\theta)\right|^2.
\end{equation}
Here $J_{0,j} = J_{0,j}^{({\rm i})}/n_{\rm s}$, the angles $\theta$ and $\theta^\prime$ are related by Snell's law
$\sin\theta = n_{\rm s}\sin\theta^\prime$, and $A_j(\theta) = t_j(\theta,\theta^\prime)A_j(\theta^\prime)$, where $t_j(\theta,\theta^\prime)$ are given by Fresnel's equations
\begin{equation}
t_1(\theta,\theta^\prime) = \frac{2n_{\rm s}\cos\theta}{\cos\theta^\prime + n_{\rm s}\cos\theta},
\end{equation}
\begin{equation}
t_2(\theta,\theta^\prime) = \frac{2\cos\theta}{n_{\rm s}\cos\theta^\prime + \cos\theta}
\end{equation}
for radial (TM) and azimuthal (TE) polarizations.

Because of the large refractive index $n_{\rm s}$ of a semiconductor, only a narrow cone of plane waves
with incident angles $\theta^\prime$ in the range $0\leq \theta^\prime < \arcsin\left(1/n_{\rm s}\right)$.
We may thus assume that the radial and azimuthal contributions to the radiant intensity of the primary source
are equal, i.e., $I_1^{({\rm i})} = I_2^{({\rm i})}$, and hence we may denote $J_0^{(\rm i)}=2J_{0,j}^{(\rm i)}$ and $J_0=2J_{0,j}$. Then the degree of polarization given by Eq.~\eqref{Idegpol}
is
\begin{equation}
P(\theta) = \frac{\left|t_1(\theta,\theta^\prime)\right|^2 - \left|t_2(\theta,\theta^\prime)\right|^2}{\left|t_1(\theta,\theta^\prime)\right|^2 + \left|t_2(\theta,\theta^\prime)\right|^2}
\end{equation}
and the radiant intensity transforms at the interface according to
\begin{equation}
\frac{J(\theta)}{J^{({\rm i})}(\theta)} = \frac{1}{2n_{\rm s}}\frac{\cos^2\theta}{\cos^2\theta^\prime}\left[\left|t_1(\theta,\theta^\prime)\right|^2+\left|t_2(\theta,\theta^\prime)\right|^2\right].
\end{equation}
If we assume that the radiation pattern produced by the primary source is Lambertian, with
\begin{equation}
J^{({\rm i})}_j(\theta^\prime) = \frac{1}{2} J_0^{({\rm i})}\cos\theta^\prime,
\end{equation}
the radial and azimuthal contributions to the radiant intensity of the secondary source become
\begin{equation}
J_j(\theta) = \frac{1}{2} J_0\frac{\cos^2\theta}{\cos\theta^\prime}\left|t_j(\theta,\theta^\prime)\right|^2.
\end{equation}
These contributions $J_1(\theta)$ and $J_2(\theta)$ are shown by the dotted and dashed lines,
respectively, in Fig.~\ref{f:LEDgeom}b, where we have taken $n_{\rm
{\rm s}} = 3.5$. The curves fit well the $\cos^n\theta$: we obtain $n\approx 3.4 = b$
for the azimuthally polarized contribution, $n\approx 2.4 = a$ for the radially polarized
contribution, and $n\approx 2.9$ for an equally weighted sum of the two contributions.
Thus the elementary electric-field modes given by Eq.~\eqref{spcosmodes} provide good
approximations of the modes of the structure in Fig.~\ref{f:LEDgeom}a.

The degree of polarization, also plotted in Fig.~\ref{f:LEDgeom}b, increases
from a zero on-axis value (unpolarized radiation in the paraxial
domain) to $P(\theta)\approx 0.85$ when $\theta\rightarrow \pi/2$,
indicating partially polarized radiation in the non-paraxial domain.

\section{Final remarks}
\label{s:discussion}

The electromagnetic elementary-mode decomposition presented in this paper should prove useful in
in optical system modeling by field tracing methods. To this end, it is necessary to determine
elementary field modes and the weight functions of the source. The example presented above illustrates
the possibility of doing this if there is sufficient a priori information about the structure of
the source. If, however, the source properties are not known, it is necessary to determine the modal
decomposition experimentally. As in the scalar case~\cite{Vahimaa}, this can in principle be accomplished
by far-field measurements. In general, the polarization basis vectors $\mathbf{f}_j$ and the eigenvalues
$I_j$ can be determined from the polarization matrix $\boldmatrix{A}(\boldsymbol{\kappa}, \boldsymbol{\kappa})$.
This matrix can be determined by measuring, e.g., the angular dependence of the Stokes parameters of the field
in the far zone. Thus only single-point measurements across the radiation pattern are needed. Determination
of the weight functions requires, in general, two-point correlation measurements in the far zone, but this
is avoided if the field is known to be quasihomogeneous.

\section*{Acknowledgments}

This work was supported by the Academy of Finland (118951, 129155,
and 209806).

\appendix\section{Derivation of Eq.~(47)}
\label{A1}

In this Appendix we present some details of the derivation of
Eq.~(\ref{Cylinder}), which is the general representation of the
elementary electric-field modes. Denoting the circular cylindrical
coordinates by $(\rho,\phi,z)$ in the position-vector space and by
$(\kappa,\psi,z)$ in the wave vector space, and the spherical polar
coordinates by $(k,\theta,\psi)$ in the wave vector space, we have
the following relations between the unit vectors of these systems
and the Cartesian coordinates (see Fig.~\ref{f:notations}):
\begin{equation}
\begin{bmatrix}
\boldsymbol{\hat{\rho}}\\
\boldsymbol{\hat{\phi}}
\end{bmatrix}
=\mathbf{R}(\phi)
\begin{bmatrix}
\mathbf{\hat{x}}\\
\mathbf{\hat{y}}
\end{bmatrix},
\label{transform1}
\end{equation}
\begin{equation}
\begin{bmatrix}
\boldsymbol{\hat{\kappa}}\\
\boldsymbol{\hat{\psi}}
\end{bmatrix}
=\mathbf{R}(\psi)
\begin{bmatrix}
\mathbf{\hat{x}}\\
\mathbf{\hat{y}}
\end{bmatrix},
\label{transform2}
\end{equation}
\begin{equation}
\begin{bmatrix}
\mathbf{\hat{k}}\\
\boldsymbol{\hat{\theta}}
\end{bmatrix}
= \mathbf{R}(\theta)
\begin{bmatrix}
\mathbf{\hat{z}}\\
\boldsymbol{\hat{\kappa}}
\end{bmatrix},
\label{transform3}
\end{equation}
where
\begin{equation}
\mathbf{R}(\xi) =\begin{bmatrix}
\cos\xi&\sin\xi\\
-\sin\xi&\cos\xi
\end{bmatrix}
\end{equation}
is the rotation matrix and $\xi$ may stand for $\phi$, $\psi$, or
$\theta$. Using Eq.~\eqref{transform3} and recognizing that $k_z =
k\cos\theta$ and $\kappa = k \sin\theta$, we can write
Eq.~\eqref{GeneralSphericalAzimuthal} in circular cylindrical
coordinates:
\begin{equation}
\mathbf{f}_j(\boldsymbol{\kappa})= \cos\theta
f_{j,\theta}(\boldsymbol{\kappa})\boldsymbol{\hat{\kappa}}+
f_{j,\psi}(\boldsymbol{\kappa})\boldsymbol{\hat{\psi}} -\sin\theta
f_{j,\theta}(\boldsymbol{\kappa})\mathbf{\hat{z}} .
\end{equation}
Using Eqs.~\eqref{transform1} and \eqref{transform2} one can
establish the relation
\begin{equation}
\begin{bmatrix}
\boldsymbol{\hat{\kappa}}\\
\boldsymbol{\hat{\psi}}
\end{bmatrix}
=\frac{1}{2}
\begin{bmatrix}
1 & 1 \\
-{\rm i} & {\rm i}
\end{bmatrix}
\begin{bmatrix}
{\rm e}^{{\rm i}\beta}\left(\boldsymbol{\hat{\rho}} + {\rm i}\boldsymbol{\hat{\psi}}\right)\\
{\rm e}^{-{\rm i}\beta}\left(\boldsymbol{\hat{\rho}} - {\rm
i}\boldsymbol{\hat{\psi}}\right)
\end{bmatrix},
\end{equation}
where $\beta = \phi -\psi$, and express
$\mathbf{f}_j(\boldsymbol{\kappa})$ in the form
\begin{eqnarray}
\mathbf{f}_j(\theta,\psi) & = & \frac{1}{2}\left[\cos\theta
f_{j,\theta}(\theta,\psi)-{\rm i} f_{j,\psi}(\theta,\psi)\right]{\rm
e}^{{\rm i}\beta}\left(\boldsymbol{\hat{\rho}} + {\rm
i}\boldsymbol{\hat{\psi}}\right)\nonumber\\
& + & \frac{1}{2}\left[\cos\theta f_{j,\theta}(\theta,\psi)+{\rm i}
f_{j,\psi}(\theta,\psi)\right]{\rm e}^{-{\rm
i}\beta}\left(\boldsymbol{\hat{\rho}} - {\rm
i}\boldsymbol{\hat{\psi}}\right)\nonumber\\
& - & \sin\theta f_{j,\theta}(\theta,\psi)\mathbf{\hat{z}}.
\label{a6}
\end{eqnarray}
Recognizing further that $\mathbf{k}\cdot \mathbf{r} =
\boldsymbol{\kappa}\cdot \boldsymbol{\rho} + k_zz$ and that
$\boldsymbol{\kappa}\cdot \boldsymbol{\rho} =
k\rho\sin\theta\cos\beta$, we can expand the plane-wave term in
Eq.~\eqref{IFT-p} in a series form
\begin{equation}
\exp\left({\rm i}\mathbf{k}\cdot\mathbf{r}\right) = \exp\left({\rm
i}kz\cos\theta\right) \sum_{m=-\infty}^\infty {\rm i}^m
J_m(k\rho\sin\theta){\rm e}^{-{\rm i}m\beta},
\end{equation}
where $J_m$ denotes the Bessel function of the first kind and order
$m$. Combining Eqs.~\eqref{AngspecG} and \eqref{a6} we then have
\begin{eqnarray}
\mathbf{e}_j(\mathbf{r}) & = & \frac{k}{2\pi}\sum_{m=-\infty}^\infty
{\rm i}^m \int_0^{\pi/2}\mathbf{g}_{j,m}(\mathbf{r})\exp\left({\rm
i}kz\cos\theta\right)\nonumber\\
& & \times J_m(k\rho\sin\theta)k\sin\theta\,{\rm
d}\theta,\label{ejr}
\end{eqnarray}
where
\begin{equation}
\mathbf{g}_{j,m}(\mathbf{r}) =
\frac{1}{2\pi}\int_0^{2\pi}\mathbf{f}_{j}(\theta,\psi){\rm e}^{-{\rm
i}m\beta}{\rm d}\psi.
\end{equation}
Performing the integration with respect to $\psi$ with the aid of
the definition of the azimuthal Fourier coefficients in
Eq.~\eqref{fouriercoeff}, we obtain
\begin{eqnarray}
\mathbf{g}_{j,m}(\mathbf{r}) & = &
\frac{1}{2}\left(\boldsymbol{\hat{\rho}} + {\rm
i}\boldsymbol{\hat{\phi}}\right)\left[\cos\theta
f_{j,\theta,m+1}(\theta)-{\rm
i}f_{j,\psi,m+1}(\theta)\right]\nonumber\\
& & \times {\rm e}^{{\rm
i}(m+1)\phi} J_m(k\rho\sin\theta)k\sin\theta\nonumber\\
& + & \frac{1}{2}\left(\boldsymbol{\hat{\rho}} - {\rm
i}\boldsymbol{\hat{\phi}}\right)\left[\cos\theta
f_{j,\theta,m-1}(\theta)+{\rm
i}f_{j,\psi,m-1}(\theta)\right]\nonumber\\
& & \times {\rm e}^{{\rm
i}(m-1)\phi}J_m(k\rho\sin\theta)k\sin\theta\nonumber\\
& - & \mathbf{\hat{z}}\,{\rm e}^{{\rm i}m\phi}
J_m(k\rho\sin\theta)k\sin^2\theta.
\end{eqnarray}
Collecting terms in the summation of Eq.~\eqref{ejr} by replacements
\begin{equation}
{\rm e}^{{\rm i}(m\pm 1)\phi}J_m(k\rho\sin\theta)\rightarrow {\rm
i}^{\mp 1}{\rm e}^{{\rm i}m\phi}J_{m\mp 1}(k\rho\sin\theta)
\end{equation}
and employing the Bessel-function identities~\cite{Arfken}
\begin{equation}
J_{m-1}(x) + J_{m+1}(x)=\frac{2m}{x}J_{m}(x)
\end{equation}
and
\begin{equation}
J_{m-1}(x)-J_{m+1}(x) = 2\frac{\mathrm d}{\mathrm{d}x}J_{m}(x)
\label{BesselDiff}
\end{equation}
we finally arrive at Eq.~\eqref{Cylinder}.

\section{Derivation of an integral formula} \label{A2}

Let us first recall the series-representation of Bessel
functions~\cite{Arfken}:
\begin{equation}
J_m(\gamma)
=
\sum_{j=0}^\infty \frac{(-1)^j}{j!(m+j)!}\left(\frac{\gamma}{2}\right)^{m+2j}.
\label{BesselSeries}
\end{equation}
On the other hand, we have the relation~\cite{Gradshteyn}
\begin{align}
\int_0^{\pi/2} \sin^p\theta\cos^q\theta\,\mathrm{d}\theta
=\frac{1}{2}B\left(\frac{p+1}{2},\frac{q+1}{2}\right),
\label{SinCosIntegral}
\end{align}
where $\Re(p)>-1$, $\Re(q)>-1$,
\begin{equation}
B(\gamma,\xi)=\frac{\Gamma(\gamma)\Gamma(\xi)}{\Gamma(\gamma+\xi)}
\end{equation}
is the beta function, and $\Gamma(\gamma)$ is the Gamma
function~\cite{Arfken}. Combining Eqs.~\eqref{BesselSeries}
and~\eqref{SinCosIntegral}, and interchanging the order of summation
and integration yields
\begin{subequations}
\begin{align}
&\int_0^{\pi/2}\sin^p\theta\cos^q\theta J_m(k\rho\sin\theta)\,\mathrm{d}\theta\nonumber\\
&=
\frac{(k\rho)^m}{2^{m+1}}\Gamma\left(\frac{q+1}{2}\right)\nonumber\\
&\quad\times
\sum_{j=0}^\infty
\frac{\Gamma[j+\frac{1}{2}(p+m+1)]}{j!\Gamma(m+j+1)\Gamma[j+1+\frac{1}{2}(p+m+q)]}\nonumber\\
&\quad\times \left(\frac{-k^2\rho^2}{4}\right)^j,
\label{IntegralSolution1}
\end{align}
where we have employed the identity $s!=\Gamma(s+1)$ for integer
$s$. Equation~\eqref{IntegralSolution1} can also be expressed in the
form
\begin{align}
&\int_0^{\pi/2}\sin^p\theta\cos^q\theta J_m(k\rho\sin\theta)\,\mathrm{d}\theta\nonumber\\
&=
\frac{(k\rho)^m}{2^{m+1}}\Gamma\left(\frac{q+1}{2}\right)\Gamma\left(\frac{p+m+1}{2}\right)\nonumber\\
&\quad\times \sideset{_1}{_2}\ER\left(
\frac{p+m+1}{2};m+1,1+\frac{p+q+m}{2};\frac{-k^2\rho^2}{4} \right),
\label{IntegralSolution2}
\end{align}
\end{subequations}
where $\sideset{_p}{_q}\ER\left(a_1,a_2,\dots,a_p;b_1,b_2,\dots,b_q;\gamma\right)$ denotes the regularized hypergeometric function, defined by
\begin{align}
&\sideset{_p}{_q}\ER\left(a_1,a_2,\dots,a_p;b_1,b_2,\dots,b_q;\gamma\right)\nonumber\\
&=\frac{\sideset{_p}{_q}\EF\left(a_1,a_2,\dots,a_p;b_1,b_2,\dots,b_q;\gamma\right)}%
{\Gamma(b_1)\Gamma(b_2)\dots\Gamma(b_q)},
\end{align}
$\sideset{_p}{_q}\EF\left(a_1,a_2,\dots,a_p;b_1,b_2,\dots,b_q;\gamma\right)$ is the generalized hypergeometric function~\cite{Bailey}
\begin{align}
&\sideset{_p}{_q}\EF\left(a_1,a_2,\dots,a_p;b_1,b_2,\dots,b_q;\gamma\right)\nonumber\\
&=\sum_{j=0}^\infty\frac{(a_1)_j(a_2)_j\dots(a_p)_j\gamma^j}%
{j!(b_1)_j(b_2)_j\dots(b_q)_j}
\end{align}
and
\begin{equation}
(c)_j=\frac{\Gamma(j+c)}{\Gamma(c)}.
\end{equation}


\begin{thebibliography}{10.}
\bibitem{Mandel}
L. Mandel and E. Wolf, \emph{Optical Coherence and Quantum Optics}
(Cambridge University Press, Cambridge, UK, 1995).
\bibitem{Wolf82}
E. Wolf, ``New theory of partial coherence in the space--frequency
domain. Part I: Spectra and cross spectra of steady-state sources,''
J. Opt. Soc. Am. {\bf 72,} 343--351 (1982).
\bibitem{Gori03}
F. Gori, M. Santarsiero, R. Simon, G. Piquero, R. Borghi, and G. Guattari,
``Coherent-mode decomposition of partially coherent, partially polarized sources,''
J. Opt. Soc. Am. A {\bf 20,} 78--84 (2003).
\bibitem{Tervo04}
J. Tervo, T. Set\"al\"a, and A. T. Friberg, ``Theory of partially coherent
electromagnetic fields in the space--frequency domain,'' J. Opt. Soc. Am. A {\bf 21,}
2205--2215 (2004).
\bibitem{Gori80a}
F. Gori, ``Collett-Wolf sources and multimode lasers,'' Opt. Commun. {\bf 34,}
301--305 (1980).
\bibitem{Gori80b}
F. Gori, ``Mode propagation of the light field generated by Collett-Wolf Schell-model
sources,'' Opt. Commun. {\bf 46,} 149--154 (1983).
\bibitem{Starikov1}
A. Starikov and E. Wolf, ``Coherent-mode representation of Gaussian Schell-model sources
and of their radiation fields,'' J. Opt. Soc. Am. A {\bf 72,} 923--928 (1982).
\bibitem{Starikov2}
A. Starikov, ``Effective number of degrees of freedom of partially coherent sources,''
J. Opt. Soc. Am. {\bf 73,} 1538--1544 (1983).
\bibitem{Huttunen}
J. Huttunen, A. T. Friberg, and J. Turunen, ``Diffraction of partially coherent electromagnetic fields
by microstructured media,'' Phys. Rev. E \textbf{52,} 3081--3092 (1995).
\bibitem{Vahimaa97}
P. Vahimaa and J. Turunen, ``Bragg diffraction of
spatially partially coherent
fields,'' J. Opt. Soc. Am. A \textbf{14,} 54--59 (1997).
\bibitem{Gori78}
F. Gori and C. Palma, ``Partially coherent sources which give rise
to highly directional laser beams,'' Opt. Commun. {\bf 27,} 185--188
(1978).
\bibitem{Gori80}
F. Gori, ``Directionality and partial coherence,'' Opt. Acta {\bf
27,} 1025--1034 (1980).
\bibitem{Vahimaa}
P. Vahimaa and J. Turunen, ``Finite-elementary-source model for
partially coherent radiation,'' Opt. Express {\bf 14,} 1376--1381
(2006).
\bibitem{Gori07}
F. Gori and M. Santarsiero, ``Devising genuine correlation
functions,'' Opt. Lett. {\bf 32,} 531-3533 (2007).
\bibitem{Turunen08}
J. Turunen and P. Vahimaa, ``Independent-elementary-field
model for three-dimensional spatially partially coherent
sources,'' Opt. Express {\bf 16,} 6433--6442 (2008).
\bibitem{Alonso08}
M. A. Alonso and E. Wolf, ``The cross-spectral density matrix of a
planar, electromagnetic stochastic source as a correlation matrix,''
Opt. Commun. \textbf{281,} 2393--2396 (2008).
\bibitem{Tervo02}
J. Tervo and J. Turunen, ``Angular spectrum representation of
partially coherent electromagnetic fields,'' Opt. Commun.
\textbf{209,} 7--16 (2002).
\bibitem{Setala2004}
T. Set\"al\"a, J. Tervo, and A. T. Friberg, ``Complete
electromagnetic coherence in space--frequency domain,'' Opt. Lett.
{\bf 29,} 328--330 (2004).
\bibitem{Tervo2003}
J. Tervo, T. Set\"al\"a, and A. T. Friberg, ``Degree of coherence
for electromagnetic fields,'' Opt. Express {\bf 11,} 1137--1143
(2003).
\bibitem{Arfken}
G. B. Arfken and H. J. Weber, \emph{Mathematical Methods for
Physicists}, 5th ed. (Academic Press, San Diego, CA, 2001).
\bibitem{Gradshteyn}
I. S. Gradshteyn and I. M. Ryzhik, {\em Table of Integrals, Series, and Products},
A. Jeffrey, D. Zwillinger, eds., seventh edition. (Academic Press, Orlando, 2007).
\bibitem{Bailey}
W. N. Bailey, \emph{Generalized Hypergeometric Series}, Vol.\ 32 of
Cambridge Tracts in Mathematics and Mathematical Physics, G. H.
Hardy and E. Cunningham, eds. (Cambridge University Press,
Cambridge, UK, 1935).

\end{thebibliography}
\end{document}